\documentclass[preprint,epsf,preprintnumbers,nofootinbib]{revtex4-1}
\usepackage{graphicx}
\usepackage{dcolumn}
\usepackage{bm}
\usepackage{amsmath}
\usepackage{amsfonts} 
\usepackage{latexsym}
\usepackage{bbm}
\usepackage{color}
\newcommand{\sumint}{\mbox{$\sum$}\kern-2.7ex\int}

\begin{document}

\preprint{NCTS-PH/1723}
\preprint{IPMU17-0109}

\title{On gauge dependence of gravitational waves from a first-order phase transition
in classical scale-invariant $U(1)'$ models}

\author{Cheng-Wei Chiang$^{1,2,3,4}$}%
\email{chengwei@phys.ntu.edu.tw}
\author{Eibun Senaha$^{1,5}$}%
\email{senaha@ibs.re.kr}

\affiliation{$^1$Department of Physics, National Taiwan University, Taipei 10617, Taiwan}
\affiliation{$^2$Institute of Physics, Academia Sinica, Taipei 11529, Taiwan}
\affiliation{$^3$Physics Division, National Center for Theoretical Sciences, Hsinchu 30013, Taiwan}
\affiliation{$^4$Kavli IPMU, University of Tokyo, Kashiwa, 277-8583, Japan}
\affiliation{$^5$Center for Theoretical Physics of the Universe, Institute for Basic Science (IBS), Daejeon 34051, Korea}

\bigskip

\date{\today}

\begin{abstract}
We study gauge dependence of gravitational waves produced from a first-order phase transition
in classical scale-invariant $U(1)'$ models. 
Accidental gauge independence of the one-loop effective potential in this class of models is
spoiled by including thermal resummation.
The gauge artifact in the resummed effective potential propagates to the gravitational wave spectrum and results in one order of magnitude uncertainties in the prediction under a specific gauge choice.
\end{abstract}


\maketitle

Existence of gravitational waves (GWs) has been confirmed by the LIGO experiment~\cite{LIGOdiscovery},
opening the door to a new era of observational astrophysics and cosmology. 
In particular, probing GWs from the early Universe may unveil the thermal history 
of the Universe as the GWs may be produced when it undergoes a first-order phase transition~\cite{GW_1stPT,GW_1stPT_EW,GW_1stPT_aboveEW,GW_1stPT_for,Grojean:2006bp,Huber:2008hg,Hindmarsh:2013xza,Caprini:2015zlo}. Therefore, a reliable prediction of the GW spectrum becomes very important.

It is a common practice to use a finite-temperature effective potential to investigate thermal phase transitions.  As is widely known, one of the thorny issues in such analyses is that the effective potential has a dependence on the gauge-fixing parameter, $\xi$~\cite{Jackiw:1974cv}.
According to the Nielsen-Fukuda-Kugo (NFK) identities~\cite{NFK}, only energies at stationary points are free from the $\xi$ dependence.
Nevertheless, the statement is not so obvious when one uses the effective potential 
in perturbative calculations. 
For example, the minimum of the one-loop effective potential still has a dependence on $\xi$,
except at the point that minimizes the {\it tree-level} potential (for an illuminating discussion, see Ref.~\cite{Patel:2011th} and the references therein).  
Therefore, the gauge artifact in the standard perturbative treatment of the effective potential 
could propagate to the predicted GW spectrum even though physical quantities should not depend on the choice of $\xi$.

The $\xi$ dependence of GWs in a massive Abelian Higgs model was studied in Ref.~\cite{Wainwright:2011qy},
which pointed out that the peak frequency in the GW spectrum could change by several orders of magnitude 
when varying $\xi$ from 0 to 5, with details depending on the input parameters.
It was also found that the results in the Landau gauge ($\xi=0$) were close to those obtained using a gauge-invariant Hamiltonian formalism~\cite{Boyanovsky:1996dc}. 
Nevertheless, as the thermal resummation has not been implemented in the gauge-invariant formalism,
the gauge-dependence issue is not yet settled, as emphasized in Ref.~\cite{Wainwright:2011qy}.
\footnote{A different observation is made in Ref.~\cite{Garny:2012cg}.}

Much attention has been paid to the Standard Model (SM) with an extra local $U(1)$ symmetry in the context of grand unification constructions
(for a comprehensive review, see Ref.~\cite{Langacker:2008yv} and references therein) and/or 
phenomenological motivations such as a solution to experimental anomalies~\cite{Crivellin:2015mga}.
Some models may have the GWs associated with the first-order $U(1)'$ transition. 
As in the simple Abelian Higgs model, the GW spectrum in those models would also suffer from the significant gauge artifact,
and thus the numerical assessment of the predictions must be taken with caution.

It is worth performing a similar analysis in classical scale-invariant $U(1)'$ versions~\cite{BLSM_Iso,Jinno:2016knw,Guo:2015lxa,Das:2016zue}
that can offer an alternative solution to the gauge hierarchy problem other than supersymmetric theories, 
as inspired by a Bardeen's naturalness argument by use of the classical scale symmetry~\cite{Bardeen:1995kv}.
The point is that once the quadratic divergence is removed by subtraction at an ultraviolet (UV) energy scale, 
it is no longer operative in the infrared (IR) regime.
This can be viewed from the renormalization group equation of the Higgs bare mass $(\mu^2)$. 
One can show that if $\mu^2=0$ at the UV energy scale, 
it remains zero in the IR regime as well due to the multiplicative renormalization property.
In this view, the quadratic divergence problem should be coped with the UV physics rather than IR (see also Refs.~\cite{SIModels_2}).
In Ref.~\cite{BLSM_Iso}, $\mu^2=0$ is imposed at the Planck scale as a boundary condition by invoking the classical scale invariance.
Moreover, an intermediate energy scale ({\it e.g.}, grand unification scale) is assumed to be absent in order not to generate a large mass 
correction from that scale. 

As discussed in Ref.~\cite{Jackiw:1974cv}, a feature of the classical scale-invariant theories is that the $\xi$-dependent terms start to show up at the two-loop order, while the one-loop effective potential remains gauge-independent by accident.  At finite temperatures, however, thermal resummation spoils the latter property and renders perturbative analyses of GW signals gauge-dependent as well.
In this regard, the numerical impact of the gauge dependence in this class of models could be potentially different from those studied in Ref.~\cite{Wainwright:2011qy}.

Even though a gauge-invariant formalism with the thermal resummation is still unknown, it is useful to estimate to what extent the GW spectrum is sensitive to the gauge choice when using a realistic parameter set in the common formalism.  In this Letter, we examine the impacts of the $\xi$ parameter on the strength of the cosmological phase transition and the spectrum of GWs generated from bubble dynamics in classical scale-invariant Abelian extensions of the SM.  As an explicit example, we present a numerical study for the $U(1)_{B-L}$ version~\cite{BLSM_Iso,Jinno:2016knw}.

We start by considering a model that is invariant under not only the SM gauge group but also extra gauged $U(1)'$ and scale symmetries.  We introduce a complex scalar field $S$ charged under the $U(1)'$ symmetry but singlet under the SM gauge group.  When $S$ spontaneously develops a vacuum expectation value (VEV), $\langle S \rangle = v_S/\sqrt2$, the $Z'$ boson associated with $U(1)'$ acquires its mass $m_{Z'} = g' Q'_S v_S$, with $g'$ and $Q'_S$ being the gauge coupling constant and the charge of $S$ associated with the $U(1)'$.  Therefore, the Lagrangian
\begin{align}
\mathcal{L} = \mathcal{L}_{\text{SM}'} - \frac{1}{4}Z'_{\mu\nu}Z'^{\mu\nu} + |D_\mu S|^2 - V(H,S) ~,
\end{align}
where $\mathcal{L}_{\text{SM}'}$ denotes the SM Lagrangian without the Higgs potential, the field strength $Z'_{\mu\nu} = \partial_\mu Z'_\nu-\partial_\nu Z'_\mu$, $D_\mu S = (\partial_\mu + ig'Q'_S Z'_\mu)S$, and $H$ denotes the $SU(2)_L$-doublet Higgs field.  The scale symmetry demands that the scalar potential be composed of only quartic interactions and read
\begin{align}
V(H, S) &= \lambda_H(H^\dagger H)^2+\lambda_{HS}H^\dagger H |S|^2+\lambda_S|S|^4 ~.
\end{align}
We parametrize $S$ as
\begin{align}
S(x) = \frac{1}{\sqrt{2}}\big(v_S+h_S(x)+iG(x)\big) ~,
\end{align}
where $G(x)$ the Nambu-Goldstone (NG) boson associated with the spontaneous breaking of $U(1)'$.
If $\lambda_{HS}$ is negative, the corresponding term in $V(H,S)$ will trigger the electroweak symmetry breaking, and result in the SM-like Higgs mass given by $m_h^2=-\lambda_{HS}v_S^2=2\lambda_Hv^2$ with $v\simeq 246$~GeV.
Here we consider a scenario in which $v_S$ is of multi-TeV, so that $-\lambda_{HS}=m_h^2/v_S^2\simeq \mathcal{O}(10^{-3})$~\cite{BLSM_Iso,Das:2016zue,Jinno:2016knw}, and $g'=\mathcal{O}(0.1) \gg |\lambda_{HS}|$.  Hence, we can analyze the $U(1)'$ phase transition independent of the SM sector.

The gauge-fixing and FP ghost terms are given by the BRS transformation of a
gauge-fixing function, $F(x)=\partial^\mu Z'_\mu(x)-\xi g'Q'_Sv_S G(x)+\xi B(x)/2$, 
where $\xi$ is the gauge-fixing parameter and 
$B(x)$ denotes the Nakanishi-Lautrup field~\cite{NL} that plays the role of a Lagrangian multiplier
for the gauge fixing~\cite{Kugo:1981hm}.  It follows that
\begin{align}
\mathcal{L}_{\text{GF}+\text{FP}} = -\frac{1}{2\xi}\Big[\partial^\mu Z'_\mu-\xi g'Q'_Sv_SG\Big]^2
	-i\bar{c}(x)\Big[\partial^\mu\partial_\mu +\xi(g'Q'_S)^2v_S\big(v_S+h_S\big)\Big]c(x) ~,
	\label{GF-FP}
\end{align}
where $c(x)$ and $\bar{c}(x)$ are the ghost and antighost fields, respectively.

As pointed out by Coleman and Weinberg~\cite{Coleman:1973jx}, the $U(1)'$ symmetry in such theories is broken by one-loop radiative corrections given by~\cite{Jackiw:1974cv}
\begin{align}
V_{\text{CW}}(\varphi_S)
=\sum_i\frac{n_i\bar{m}^4_i}{64\pi^2}\left(\ln\frac{\bar{m}^2_i}{\bar{\mu}^2}-c_i\right),
\end{align}
where $\varphi_S$ is the classical field of $S$, $\bar{m}$ is the $\varphi_S$-dependent mass of a particle of species $i$, $n_i$ is the corresponding number of degrees of freedom, $\bar{\mu}$ is the renormalization scale, and $c_i=3/2$ for scalars and FP ghosts and $5/6$ for gauge bosons.
As recognized in Ref.~\cite{Jackiw:1974cv}, $V_{\text{CW}}$ inherently depends on the $\xi$ parameter. 
The one-loop effective potential takes the form~\cite{Wainwright:2011qy,Delaunay:2007wb}
\begin{align}
V_{\text{eff}}(\varphi_S) 
&= 
\frac{\lambda_S}{4}\varphi_S^4
+\frac{\bar{m}_S^4}{64\pi^2}\left(\ln\frac{\bar{m}_S^2}{\bar{\mu}^2}-\frac{3}{2}\right)
+3\frac{\bar{m}_{Z'}^4}{64\pi^2}\left(\ln\frac{\bar{m}_{Z'}^2}{\bar{\mu}^2}-\frac{5}{6}\right)
\nonumber\\
&\qquad~~~~~
+\frac{\bar{m}_{G,\xi}^4}{64\pi^2}\left(\ln\frac{\bar{m}_{G,\xi}^2}{\bar{\mu}^2}-\frac{3}{2}\right)
-\frac{(\xi\bar{m}_{Z'}^2)^2}{64\pi^2}\left(\ln\frac{\xi\bar{m}_{Z'}^2}{\bar{\mu}^2}-\frac{3}{2}\right),
\label{VcwU1}
\end{align}
where the field-dependent masses of $S$, $Z'$, and $G$ in the $R_\xi$ gauge are respectively given by
\begin{align}
\bar{m}^2_S = 3\lambda_S\varphi_S^2 ~,\quad  
\bar{m}^2_{Z'} = (g'Q'_S\varphi_S)^2 ~,\quad
\bar{m}^2_{G,\xi} = \bar{m}^2_G+\xi \bar{m}_{Z'}^2 ~,
\end{align}
with $\bar{m}^2_G = \lambda_S \varphi_S^2$.
Even though the $\xi$-dependent terms are partly cancelled among the gauge boson, the NG boson
and the ghosts, the $\xi$ dependence still remains at this stage.

Minimizing the one-loop effective potential in Eq.~(\ref{VcwU1}) with respect to $\varphi_S$ 
and evaluating it at $\varphi_S=v_S$, one can solve for $\lambda_S$ iteratively and obtains to the leading order that
\begin{align}
\lambda_S \simeq -\frac{3m_{Z'}^4}{16\pi^2v_S^4}\left(\ln\frac{m_{Z'}^2}{\bar{\mu}^2}-\frac{1}{3}\right),
\label{eq:lambdaS}
\end{align}
where we have dropped terms of higher order in $\lambda_S$.  This result is in stark difference from the corresponding one in $U(1)'$ models without the scale symmetry.  Putting $\lambda_S$ back to Eq.~(\ref{VcwU1}), we obtain
\begin{align}
V_{\text{eff}}(\varphi_S) 
\simeq \frac{3\bar{m}_{Z'}^4}{64\pi^2}\left(\ln\frac{\varphi_S^2}{v_S^2}-\frac{1}{2}\right),
\label{rVcwU1}
\end{align}
which shows no $\xi$ dependence. It should be emphasized that in ordinary $U(1)$ models without scale invariance, $\bar{m}_G^2$ cannot be considered as a result of one-loop effects as in the above case.  In that case, $V_{\text{eff}}(\varphi_S)$ depends on $\xi$ except at the point where $\bar{m}_G^2=0$, corresponding to the parameter set when the {\it tree-level} potential, rather than the one-loop potential, assumes its minimum.  Albeit no gauge dependence shows up in Eq.~\eqref{rVcwU1}, we will point out with an explicit example below that the $\xi$ dependence cannot be relegated to the second order in perturbation at finite temperatures due to a thermal resummation.

It is well known that at high temperatures perturbative expansions break down and require {\it thermal resummation}, {\it i.e.}, reorganizing the expansions in such a way that dominant thermal pieces are summed up to all orders. 
Following the resummation method for Abelian gauge theories presented in Refs.~\cite{Buchmuller:1992rs,Funakubo:2012qc}, the thermal masses of the longitudinal and transverse parts ($\Delta m_{L,T}$) 
of the $Z'$ boson as well as the thermal mass of $S$ are added and subtracted in the unresummed Lagrangian as
\begin{align}
\mathcal{L} 
&\to 
\left[\mathcal{L}
	+\Delta m_S^2|S|^2
	+\frac{1}{2}\Delta m_L^2 Z'^\mu L_{\mu\nu}(i\partial)Z'^\nu
	+\frac{1}{2}\Delta m_T^2 Z'^\mu T_{\mu\nu}(i\partial)Z'^\nu
\right] \nonumber\\
&\qquad~
	-\Delta m_S^2|S|^2
	-\frac{1}{2}\Delta m_L^2 Z'^\mu L_{\mu\nu}(i\partial)Z'^\nu
	-\frac{1}{2}\Delta m_T^2 Z'^\mu T_{\mu\nu}(i\partial)Z'^\nu
~,
\label{resumLag}
\end{align}
where $T_{\mu\nu}$ and $L_{\mu\nu}$ are projection tensors defined by
\begin{align}
\begin{split}
&T_{00}=T_{0i}=T_{i0}=0 ~,\quad T_{ij} = g_{ij}-\frac{k_ik_j}{-\boldsymbol{k}^2} ~,
\\
&L_{\mu\nu}=P_{\mu\nu}-T_{\mu\nu} ~, \quad P_{\mu\nu}=g_{\mu\nu}-\frac{k_\mu k_\nu}{k^2}
~,
\end{split}
\end{align}
in the rest frame of the thermal bath, where $g_{\mu\nu}=\text{diag}(1,-1,-1,-1)$ and $k^\mu$ is the 4-momentum of the $Z'$ boson.  
Note that the original Lagrangian with the added terms in the square brackets in Eq.~\eqref{resumLag} are considered as an un-perturbed tree-level part, while the subtracted terms on the second line are treated as the thermal couterterms that appear at the loop order.  We also note that gauge invariance of the Lagrangian is not spoiled by the above-mentioned procedure.

With the Lagrangian given in Eq.~(\ref{resumLag}), the resummed effective potential takes the form
\begin{align}
V_{\text{eff}}(\varphi_S; T) 
&=\frac{\bar{M}^4_{L}}{64\pi^2}\left(\ln\frac{\bar{M}^2_{L}}{\bar{\mu}^2}-\frac{3}{2}\right)
	+\frac{2\bar{M}^4_T}{64\pi^2}\left(\ln\frac{\bar{M}^2_T}{\bar{\mu}^2}-\frac{1}{2}\right)	
	-\frac{3\bar{m}_{Z'}^4}{64\pi^2}\left(\ln\frac{m_{Z'}^2}{\bar{\mu}^2}-\frac{1}{3}\right)
	\nonumber\\
&\quad+\frac{(\xi\bar{m}_{Z'}^2+\Delta m_S^2)^2}{64\pi^2}\left(\ln\frac{\xi\bar{m}_{Z'}^2+\Delta m_S^2}{\bar{\mu}^2}-\frac{3}{2}\right) 
-\frac{(\xi\bar{m}_{Z'}^2)^2}{64\pi^2}\left(\ln\frac{\xi\bar{m}_{Z'}^2}{\bar{\mu}^2}-\frac{3}{2}\right)
	\nonumber\\
&\quad 
	+\frac{T^4}{2\pi^2}
	\left[
	I_B\left(\frac{\bar{M}^2_L}{T^2}\right)
	+2I_B\left(\frac{\bar{M}^2_T}{T^2}\right)
	+I_B\left(\frac{\xi\bar{m}_{Z'}^2+\Delta m_S^2}{T^2}\right) 
	-I_B\left(\frac{\xi\bar{m}_{Z'}^2}{T^2}\right) 
	\right]
	~,
\label{ResumVeff}
\end{align}
where
\begin{align}
I_B(a^2) = \int_0^\infty dx~x^2\ln\Big[1-e^{-\sqrt{x^2+a^2}}\Big]
~,
\end{align}
with $\bar{M}^2_L=\bar{m}^2_{Z'}+\Delta m_L^2$ and $\bar{M}^2_T = \bar{m}^2_{Z'}+\Delta m_T^2$.
To the leading order in high-temperature expansions, one has 
\begin{align}
\Delta m_L^2 = \frac{(g'Q'_S)^2}{3}T^2 ~,\quad
\Delta m_T^2=0 ~,\quad
\Delta m_S^2=\frac{(g'Q'_S)^2}{4}T^2 ~,
\end{align}
that are $\xi$-independent.
Note that the resummed  effective potential in Eq.~(\ref{ResumVeff}) is no longer $\xi$-independent because $\Delta m_S^2\neq 0$.
Again, we will quantify how sensitive the first-order phase transition strength and the GW spectrum are to the gauge-fixing parameter $\xi$ using an explicit model.

After the thermal resummation, one cannot completely gauge away the kinetic energy of the gauge field.
However, since such an energy is gauge-independent, we will neglect it in the following discussions for simplicity.
Furthermore, the critical bubble for the first-order phase transition in the early Universe is assumed to be spherically symmetric, with the energy functional given by
\begin{align}
S_3 
= 4\pi\int^\infty_0dr~r^2
\bigg[ \frac{1}{2}\bigg(\frac{d\phi_S}{dr}\bigg)^2 + V_{\rm eff}(\phi_S;T) \bigg]
~,
\end{align}
where $\phi_S(r)=\sqrt{2}\langle S(r)\rangle$.
The equation of motion for $\phi_S$ is then 
\begin{align}
\frac{d^2\phi_S}{dr^2}+\frac{2}{r}\frac{d\phi_S}{dr}-\frac{\partial V_{\rm eff}}{\partial \phi_S}=0
~, 
\label{EOM_CB}
\end{align}
with the boundary conditions: $\lim_{r\to\infty}\phi_S(r)=0$ and $d\phi_S(r)/dr|_{r=0}=0$.
We can solve Eq.~(\ref{EOM_CB}) by use of a relaxation method (see, {\it e.g}., 
Ref.~\cite{Funakubo:2009eg} for details).

Let $T_*$ be the temperature at which the GWs are produced from the cosmological phase transition. 
Without significant reheating, this temperature can be approximated by the bubble nucleation temperature, $T_N$, to be defined below. 
For the phase transition to develop, at least one bubble must nucleate within the Hubble volume. 
We thus define $T_N$ through the condition
\begin{align}
\Gamma_N(T_N) = H^4(T_N)
~,
\label{defTn}
\end{align}
where $H(T)=1.66\sqrt{g_*(T)}T^2/m_{\text{Pl}}$ with
$g_*(T)$ being the relativistic degrees of freedom at $T$ and
$m_{\text{Pl}}=1.22\times 10^{19}$~GeV,
while $\Gamma_N(T)$ is the bubble nucleation rate per unit time per unit volume 
approximately given by~\cite{Linde:1981zj}
\begin{align}
\Gamma_N(T) \simeq T^4\left(\frac{S_3(T)}{2\pi T}\right)^{3/2}e^{-S_3(T)/T}
~. 
\label{GamN}
\end{align}
From Eqs.~(\ref{defTn}) and  (\ref{GamN}), one obtains $S_3(T_N)/T_N\simeq 140 - 150$.

A model-independent analysis of the GWs has been done in Ref.~\cite{Grojean:2006bp}
using two parameters:
\begin{align}
\alpha \equiv \frac{\epsilon(T_*)}{\rho_{\text{rad}}(T_*)}
~~\mbox{and}~~
\beta \equiv H_*T_*\frac{d}{dT}\left(\frac{S_3(T)}{T}\right)\bigg|_{T=T_*}
~,
\end{align}
where
\begin{align}
\quad 
\epsilon(T) = \Delta V_{\text{eff}}-T\frac{\partial \Delta V_{\text{eff}}}{\partial T}
~~\mbox{and}~~
\rho_{\text{rad}}(T)=\frac{\pi^2}{30}g_*(T)T^4,
\end{align}
with $\Delta V_{\text{eff}}$ being the energy difference between the symmetric and broken 
phases, and $H_*=H(T_*)$. For notational simplicity, we also introduce $\tilde{\beta} \equiv \beta/H_*$.

During the first-order phase transition, the GWs are sourced from bubble collisions, sound waves and turbulence induced by percolation, leading to
$\Omega_{\text{GW}}h^2 = \Omega_{\text{col}}h^2+\Omega_{\text{sw}}h^2+\Omega_{\text{turb}}h^2$.
Ref.~\cite{Hindmarsh:2013xza} shows that the sound waves can be dominant around the peak frequency and its spectrum~\cite{Caprini:2015zlo}
\begin{align}
\Omega_{\text{sw}}h^2(f)	
 = 2.65\times 10^{-6}\tilde{\beta}^{-1}\left(\frac{\kappa_v\alpha}{1+\alpha}\right)^2 \left(\frac{100}{g_*}\right)^{1/3}v_w
\left(\frac{f}{f_{\text{sw}}}\right)^3\left(\frac{7}{4+3(f/f_{\text{sw}})^2}\right)^{7/2},
\label{Oh2_sw}
\end{align}
where $v_w$ denotes the bubble wall velocity, $f_{\text{sw}}$ is the peak frequency given by
\begin{align}
f_{\text{sw}} = 1.9\times 10^{-2}~\text{mHz}~\frac{\tilde{\beta}}{v_w}\left(\frac{T_*}{100~{\text GeV}}\right)
\left(\frac{g_*}{100}\right)^{1/6},
\label{fsw}
\end{align}
and $\kappa_v \simeq \alpha/(0.73+0.083\sqrt{\alpha}+\alpha)$ for $v_w\simeq 1$.
In our numerical analysis below, we will take $v_w=0.95$ as a benchmark value.
Since $\Omega_{\text{sw}}\propto f^{-4}$ while $\Omega_{\text{col}}\propto f^{-1}$
and $\Omega_{\text{turb}}\propto f^{-5/3}$~\cite{Caprini:2015zlo} at higher frequencies, 
our numerical calculations also include the other two GW sources using the formulas listed in Refs.~\cite{Huber:2008hg,Caprini:2015zlo} in order to have the correct behavior in that regime.

As an explicit example of the classical scale-invariant $U(1)'$ models, we now consider the $U(1)_{B-L}$ symmetry.  In order to be gauge anomaly-free, three right-handed neutrinos ($\nu_{R_{1,2,3}}$) are naturally introduced with the Yukawa interactions $\frac{1}{2}\sum_{i=1,2,3}Y_{\nu_{Ri}}S\bar{\nu}_{Ri}^c\nu_{Ri}+\text{H.c}$.  This implies that $Q'_S=+2$ and the right-handed neutrinos acquires Majorana mass from $v_S$ (see, {\it e.g.}, Ref.~\cite{Das:2016zue} for a detailed discussion).  Note that the singlet scalar mass at the one-loop order is given according to Eq.~(\ref{rVcwU1}) by $m_S^2 = 8Bv_S^2$, where $B=3m_{Z'}^4/(64\pi^2 v_S^4)$.  In the $U(1)_{B-L}$ case, we have $B=(3m_{Z'}^4-2\sum_{i=1,2,3}m_{\nu_{Ri}}^4)/(64\pi^2 v_S^4)$ and
from which the condition that $\sum_{i=1,2,3}m_{\nu_{Ri}}^4<3m_{Z'}^4/2$~\cite{BLSM_Iso}.  Therefore, the right-handed neutrinos cannot be arbitrarily heavy with respect to the $Z'$ mass.

To further simplify the numerical analysis without losing main features, we further suppose that the right-handed neutrinos share the same Yukawa coupling.  In this case, the model has only three new free parameters, which we choose to be $\alpha'\equiv g'^2/4\pi=0.015$, $m_{Z'}=4.5$~TeV
and $m_{\nu_{R1}}=m_{\nu_{R2}}=m_{\nu_{R3}}=1.0$~TeV,
leading to $m_S\simeq 0.76$~TeV.  This parameter choice is consistent with the recent LHC Run-II data
and perturbativity up to the Planck scale~\cite{SIZp_LHC}. 
The original parameters in the Lagrangian are correspondingly fixed as $g'=0.43$, 
$v_S\simeq 5.182$~TeV and $Y_{\nu_{R}}\equiv Y_{\nu_{R1}}=Y_{\nu_{R2}}=Y_{\nu_{R3}}=0.27$.
With this setup, one obtains $\Delta m_S^2=(g'^2+Y_{\nu_R}^2/8)T^2$.
Moreover, $\bar{\mu}$ in the resummed effective potential, Eq.~(\ref{ResumVeff}), in the current study is set to $v_S$.

\begin{table}[t]
\center
\begin{tabular}{|c|c|c|c|c|}
\hline
 & no resum & $\xi=0$ & $\xi=1$ & $\xi=5$ \\ \hline
$v_S(T_C)/T_C$ & $4.851/1.321=3.67$ & $4.833/1.346=3.59$ & $4.816/1.368=3.52$ & $4.695/1.348=3.48$ \\ 
$v_S(T_*)/T_*$ & $5.181/0.328=15.8$ & $5.181/0.368=14.1$ & $5.180/0.405=12.8$ & $5.163/0.490=10.5$ \\
\hline
$\alpha$ & 2.27 & 1.44 & 0.99 & 0.48 \\
$\tilde{\beta}$ & 89.4 & 97.5 & 105.4 & 135.0  \\
\hline
\end{tabular}
\caption{Various quantities obtained without the thermal resummation in contrast with those obtain using the resummed effective potential in Eq.~(\ref{ResumVeff}) with $\xi=0$, 1 and 5.
Dimensionful parameters are expressed in units of TeV.
We take $Q'_S=2$, $\alpha'=g'^2/4\pi=0.015$, $m_{Z'}=4.5$~TeV and $m_{\nu_{R1,2,3}}=1.0$~TeV.}
\label{tab}
\end{table}

In Table~\ref{tab}, some physical quantities are listed for the unresummed case and the
resummed case with $\xi=0$, 1 and 5. 
As a reference, we give a critical temperature at which the effective potential has two degenerate minima and the corresponding VEV at the temperature, denoted by $T_C$ and $v_S(T_C)$, respectively.
One can see that, as expected, the unresummed case yields a slightly stronger first-order phase transition 
than the ordinary $\xi$-dependent cases with the resummation.
It should be remarked that $v/T$ is less sensitive to $\xi$ at $T_C$ but not at $T_*$.  This fact eventually
affects $\alpha$ and $\tilde{\beta}$ significantly. 

\begin{figure}[th]
\center
\includegraphics[width=10cm]{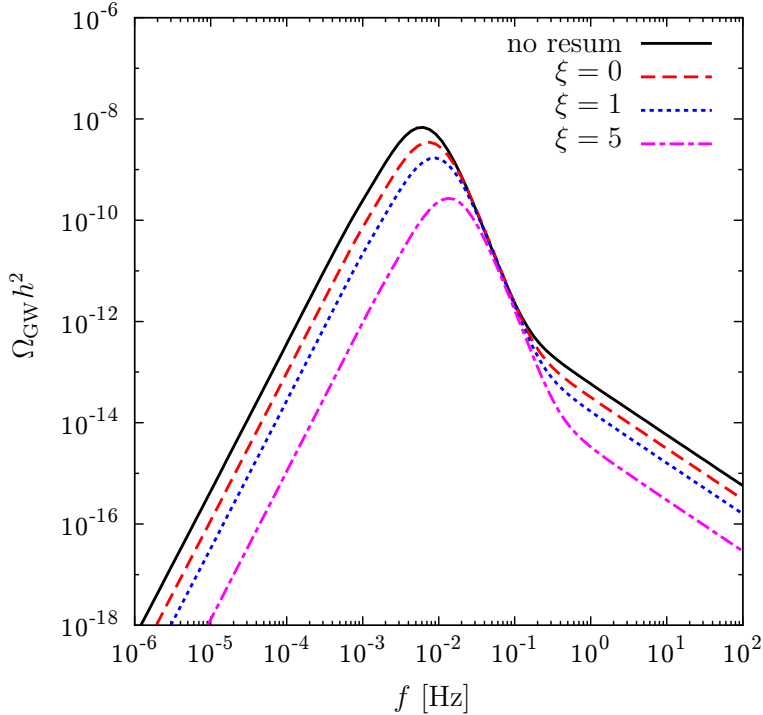}
\caption{$\Omega_{\text{GW}}h^2$ as a function of frequency.
The input parameters are the same as in Table~\ref{tab}.
The black-solid curve represents the unresummed ($\xi$-independent) case. The resummed case with $\xi=0,1$ and 5 are marked
as red-dashed, blue-dotted and magenta-dot-dashed lines, respectively.}
\label{fig}
\end{figure}

In Fig.~\ref{fig}, $\Omega_{\text{GW}}h^2$ is plotted as a function of the GW frequency $f$. 
The spectrum obtained without the thermal resummation is given by the black solid curve,
while those with the thermal resummation with $\xi=0$, 1 and 5 are plotted in red-dashed, 
blue-dotted and magenta-dot-dashed curves, respectively. 
As shown, the dependence of the GW spectrum on $\xi$ is significant, with around one order of magnitude decrease as $\xi$ changes from 0 to 5 and the peak frequency shifting toward higher frequencies.  This is primarily due to the fact that 
$\Omega_{\text{sw}}h^2\propto \tilde{\beta}^{-1}\alpha^2/(1+\alpha)^2$ and $f_{\text{sw}}\propto \tilde{\beta}$, as seen in Eqs.~(\ref{Oh2_sw}) and (\ref{fsw}).  The change in the slopes of the curves around $f \simeq 0.1$~Hz is because, as alluded to before, the GWs produced from bubble collisions and turbulence become more dominant than those from the sound waves at higher frequencies.

Depending on the input parameters $\alpha'$, $m_{Z'}$ and $m_{\nu_{R1,2,3}}$,
the strength of the first-order phase transition in the $U(1)_{B - L}$ model and the GW spectrum can change.
Nevertheless, we find the general tendency that $\Omega_{\text{GW}}h^2$ is reduced by about one order of magnitude as $\xi$ varies from 0 to 5.
We also note that there is no sensible reason why $\xi$ should restricted to the range of $[0,5]$
{\it a priori}.  We find that $\Omega_{\text{GW}}h^2$ decreases more and the peak frequency shifts higher for $\xi>5$. 
For $\xi$ larger than a certain value, however, it is found that the $U(1)'$ symmetry cannot be restored even at sufficient high temperatures in some cases
(for other unphysical artifact issues along the same line, see Ref.~\cite{Wainwright:2011qy}).
Therefore, our estimation of the sensitivity of $\Omega_{\text{GW}}h^2$ on the gauge-fixing parameter presented in this work is conservative.

In summary, we have discussed the gauge artifact in the strength of the first-order phase transition 
and the gravitational wave spectrum in the classical scale-invariant $U(1)'$ models.
We have explicitly shown that the gauge dependence re-enters the one-loop effective potential through the thermal resummation required at high temperatures.  This gauge dependence propagates to the prediction of the gravitational wave spectrum.  Through a general consideration, the significant gauge sensitivity in $\Omega_{\text{GW}}h^2$ observed in Ref.~\cite{Wainwright:2011qy} for a massive Abelian Higgs model is shown to also appear in the classical scale-invariant $U(1)'$ models.  As an explicit example of this class of models, we consider the anomaly-free $U(1)_{B - L}$ model.  As we vary the gauge-fixing parameter $\xi$ from 0 to 5 using a set of model parameters consistent with the current LHC Run-II data and perturbativity, the peak of $\Omega_{\text{GW}}h^2$ reduces by about one order of magnitude and shifts toward higher frequencies.  Such a result gives us useful information about uncertainties in the calculation of the gravitational wave spectrum done with a specific choice of gauge.  A gauge-invariant formalism for the thermal resummation is thus required for obtaining a more reliable prediction.

\begin{acknowledgments}
This work was supported in part by the Ministry of Science and Technology of Taiwan under Grant Nos. 104-2628-M-008-004-MY4 and 104-2811-M-008-056, and IBS under the project code, IBS-R018-D1 (ES). C.-W.~C would like thank the hospitality of the Theoretical Particle Physics Group at Kyoto University where part of this work was done during his visit.
\end{acknowledgments}


\end{document}